\documentclass[sigconf]{acmart}
\usepackage{multirow}
\AtBeginDocument{%
  \providecommand\BibTeX{{%
    \normalfont B\kern-0.5em{\scshape i\kern-0.25em b}\kern-0.8em\TeX}}}

\copyrightyear{2023}
\acmYear{2023}
\setcopyright{cc}
\acmConference[WWW '23]{Proceedings of the ACM Web Conference 2023}{April 30--May 4, 2023}{Austin, TX, USA}
\acmBooktitle{Proceedings of the ACM Web Conference 2023 (WWW '23), April 30--May 4, 2023, Austin, TX, USA}
\acmDOI{10.1145/3543507.3583442}
\acmISBN{978-1-4503-9416-1/23/04}

\begin{document}

\title[Migration Reframed]{Migration Reframed? A multilingual analysis on the stance shift in Europe during the Ukrainian crisis}

\author{Sergej Wildemann}
\affiliation{%
  \department{L3S Research Center}
  \institution{Leibniz Universität Hannover}
  \country{Germany}
}
\email{wildemann@l3s.uni-hannover.de}
\orcid{0000-0001-9216-4253}

\author{Claudia Nieder\'ee}
\affiliation{%
  \department{L3S Research Center}
  \institution{Leibniz Universität Hannover}
  \country{Germany}
}
\email{niederee@l3s.uni-hannover.de}
\orcid{0000-0002-2959-0484}

\author{Erick Elejalde}
\affiliation{%
  \department{L3S Research Center}
  \institution{Leibniz Universität Hannover}
  \country{Germany}
}
\email{elejalde@l3s.uni-hannover.de}
\orcid{0000-0002-4755-1606}

\renewcommand{\shortauthors}{Wildemann, et al.}

\begin{abstract}
The war in Ukraine seems to have positively changed the attitude toward the critical societal topic of migration in Europe -- at least towards refugees from Ukraine. 
We investigate whether this impression is substantiated by how the topic is reflected in online news and social media, thus linking the representation of the issue on the Web to its perception in society. 
For this purpose, we combine and adapt leading-edge automatic text processing for a novel multilingual stance detection approach. 
Starting from 5.5M Twitter posts published by 565 European news outlets in one year, beginning September 2021, plus replies, we perform a multilingual analysis of migration-related media coverage and associated social media interaction for Europe and selected European countries.

The results of our analysis show that there is actually a reframing of the discussion illustrated by the terminology change, e.g., from ``migrant'' to ``refugee'', often even accentuated with phrases such as ``real refugees''.
However, concerning a stance shift in public perception, the picture is more diverse than expected. 
All analyzed cases show a noticeable temporal stance shift around the start of the war in Ukraine.
Still, there are apparent national differences in the size and stability of this shift.  
\end{abstract}




\begin{teaserfigure}
  \includegraphics[width=\textwidth]{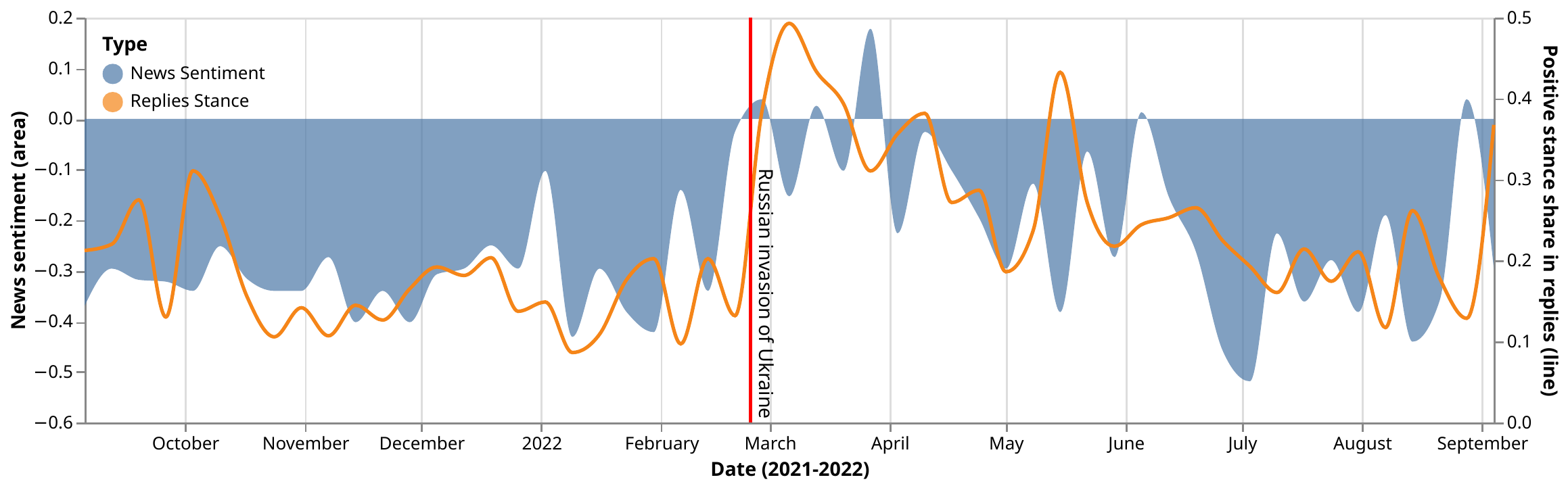}
  \caption{Weekly median news tweet sentiment and replies stance on the migration topic of 5 European countries.}
  \Description{The median online sentiment trend had a significant change after February 2022. It went from very negative (around -0.4) to mostly neutral/positive for about 4 months. The relative share of positive stance towards migrants and refugees in the replies follows the sentiment curve.}
  \label{fig:teaser}
\end{teaserfigure}

\maketitle

\section{Introduction}
The war in Ukraine has created a large surge of refugees leaving the country.
Four months after the beginning of the 2022 invasion, the UN Refugee Agency (UNHCR) reported 5.1 million\footnote{Until June 16th, there were 7.7 million border crossing from Ukraine since February, but 2.5 million crossed back to Ukraine in the same period~\cite{UNHCR_2022}.} Ukrainian refugees~\cite{UNHCR_2022}.
By September 2022, over 4M people have registered for protection schemes in European countries, especially in Poland with more than 1.3M and Germany with more than 700k persons. %
This makes the situation in sheer numbers comparable with the European refugee crisis of 2015-2018~\cite{UNHCR_2015}.   

The conflict in Ukraine and the resulting migration received a lot of attention in media and social media.
At the same time, there seems to be a more positive framing of migration, at least towards refugees from Ukraine.
Before the escalation of the conflict, media coverage and government policies on migration were mainly focused on other groups coming from conflict zones, such as Syria, Ethiopia, or Afghanistan.
However, they are often covered in connection to both economic and security threats \cite{eberl2021mapping, cooper2021beyond}. %
Attitudes towards migration are influenced by information from the press \cite{kosho2016media, dennison2018public} as well as by political agendas in the respective European countries.
Media analysis has shown frequent negative terminology such as ``illegal'', ``violence'', ``terrorist'' used in this context \cite{eberl2018european, couttenier2019logic, cooper2021beyond}.
Moreover, news reports have linked certain crimes or socioeconomic issues (e.g., the rise of unemployment) to immigration \cite{wigger2019anti, benesch2019media}. 
In contrast, in 2022, the humanitarian crisis at the EU border resulting from Ukrainians fleeing the war prompted a massive reaction of support by the western media and a great display of solidarity from the European public in particular.
Several differences might have triggered this change, including (a) the intensive reporting about the war situation raises empathy (there was also a more ``welcoming'' culture during the war in Syria) and (b) the demographic difference in the migrant population (about 90 percent of the refugees from Ukraine are female~\cite{UNHCR_2022}). %
Another significant difference seems to be (c) the cultural proximity of Ukraine to the EU compared to refugees from the Middle East \cite{therova2022anti}. 

In our work, we investigate if the impression of a stance shift towards migration in the media and the public perception in Europe can be substantiated with evidence from social media deliberation.
    To characterize the change in the tone of the discussion, we contrast the language and related sentiment used by the mass media in their online coverage of refugee crises before and after February 2022 (see Figure~\ref{fig:teaser}).
Moreover, we examine the reaction of the EU audience through their engagement on Twitter with related news.
In particular, the mass media online presence plays a prominent role in setting the discussion agenda and constitutes a reference news source for many \cite{houston2020twitterization, enli2018trust}.
Furthermore, the abundance of audience interactions on Twitter with the news provides a precious source of data for understanding users' engagement patterns and evolving opinions on sensitive topics \cite{rovetta2021influence, lazarus2021bully}.
With our work, we analyze the questions: 
Is there a stance shift with respect to migration triggered by the war in Ukraine and the reporting about it?
Are there differences between individual European countries with respect to this?
Can we show a relationship between media reporting and stance as reflected in social media reactions?

Our main contributions can be summarized as follows:
\begin{itemize}
\item we propose a novel approach for multilingual stance detection based on leading-edge technology, which combines a precise and strongly context-driven data collection with an inherently multilingual analysis process, that can generalize to further languages, thus, easing cross-country studies;
\item Based on a study of five European countries, we show a positive, at least temporal, shift of media sentiment and public stance toward the migration topic in Europe, starting with the Russia-Ukraine War -- and we also show that this shift differs in strength and stability between the countries.
\item We create a multilingual Twitter dataset consisting of $8,242$ news and reply pairs across five European countries with annotated stance toward migrants/refugees, which can serve as a valuable source for developing further analysis methods;
\end{itemize}

\section{Recent Migration to Europe}
\label{Sec:BackRel}
Terminology and language use play an important role in framing the discussion on migration.
This already starts with the terms ``migrant'' and ``refugee''.
According to UNHCR, ``\emph{Migrants} choose to move not because of a direct threat of persecution or death, but mainly to improve their lives by finding work, or in some cases for education, family reunion, or other reasons''~\cite{UNHCR_viewpoint_refugee_migrant}.
A refugee, according to the 1951 Refugee Convention~\cite{UNHCR_convention_protocol}, is ``someone who is unable or unwilling to return to their country of origin owing to a well-founded fear of being persecuted for reasons of race, religion, nationality, membership of a particular social group, or political opinion.''.
Especially during periods of crisis, media coverage defaults to biased uses of the terminologies ``refugee'' or ``migrant'' to frame the situation in their way.
The selected language is crucial, as it influences the perceived legitimacy of claims to international protection and justifies policies of exclusion and containment \cite{doi:10.1080/1369183X.2017.1348224, long2013refugees}.
This framing of the discussion based on the migrant/refugee dichotomy has also been observed in social media debates \cite{10.1145/3201064.3201087}.

The International Centre for Migration Policy Development (ICMPD) identifies the displacement of populations due to (civil) wars and conflicts as one of the main drivers of migration flow \cite{ICMPD_report_2022}. 
Impacted by decades of wars, the Middle East (especially Syria, Iraq, and Afghanistan) has some of the most significant displacement crises in the world.
Most refugees looking for asylum during the 2015 EU crisis were from the Middle East. 
This was the largest influx of migrants in a single year since World War II.
Up to 2022, over 6.8M refugees have been forced to flee Syria alone~\cite{UNHCR_syria_refugee_crisis}.
Neighboring countries such as Turkey and Lebanon have received the vast majority of them.
However, around one million Syrian refugees have moved outside the region with most hosted within the EU.  %

More recently, in 2021, the EU experienced another migration crisis with mostly Iraqi and Syrian refugees at the EU's external borders to Belarus. %
The situation did by far not reach the scale of 2015, but still received ample coverage from the mass media. %
This time, the events were reported as instigated by Belarusian authorities and politically motivated\footnote{\url{https://www.bbc.com/news/world-europe-59237413}}.
Still, the drivers for individual immigrants from these countries continued to be local hostilities and violence re-intensified throughout 2021.

In this topic, the 2022 crisis in Ukraine and the media reception of people fleeing after the Russian invasion brought attention to a potential differential treatment of other asylum seekers.
The contrast was quite prominent in Poland, one of the most affected countries by various crises within a few months\footnote{\url{https://www.politico.eu/article/poland-two-very-different-borders-ukraine-belarus-war-refugees/}}.
It is also worth mentioning that before 2022, Ukraine was already a relevant migration country of origin, with 6.1M migrants worldwide in 2020 \cite{UN_migrant_stock_2020}. %

\section{Related Work}
Previous studies on migration have leveraged details from political and media campaigns to understand changes in public attitudes \cite{barna2019attitude, valentino2013immigration}.
The theory suggests that there are two main antecedents for migration opinion.
On the one hand, we have the discourse around the issue of self-interest, and on the other hand, the role of social identity (ethnocentrism) \cite{therova2022anti}.
Media framing of these issues is very relevant among the factors influencing public opinion \cite{kosho2016media, koch2020public, giorgi2017migrants, dennison2018public}. 
In particular, politicization and securitization of the topic in Europe have been proven to shape public attitudes, as news organizations and political parties link migration to security issue \cite{gattinara2017politicization}.
The intensity of the perceived threat seems higher when covered by right-wing press \cite{troszynski2022great}.
Specifically for Polish media discourse on migration, previous studies already show a significant polarization on the topic, especially along the lines of conservative and liberal media \cite{troszynski2022great}.
However, evidence suggests that consensus is more likely to emerge when the discussion focuses on specific types of migration  \cite{valentino2013immigration, dennison2018public}.
We expect to see this effect around significant events or specific migration crises, where the news coverage about immigration may concentrate on a particular group.

Nevertheless, as the traditional role of news outlets shifts and social media becomes the primary source of news for a growing part of the western public \cite{pew_news_impact_2021,ORELLANARODRIGUEZ201874}, more evidence is required to understand the relation between media coverage and public opinion.

Research on public opinion and perception of immigrants has been traditionally conducted through surveys \cite{polish_perception_migration, therova2022anti}.
However, nowadays, social media platforms like Twitter play a significant role in public debates in many countries \cite{highfield2017social, sujon2021social}.
In recent years, social media data has supported multiple studies of several socio-cultural and political phenomena.
Some examples include political stance \cite{polquiz_politicalbias, kubin2021role}, the association of religion with users' happiness \cite{ritter2014happy}, and, more recently, opinion polarization during the COVID-19 pandemic \cite{jiang2021polarization}.
In particular, for the subject of migration, social media provide a proxy to monitor public interest and perception~\cite{khatua2022unraveling, ozturk2018sentiment, freire2021framework, khatua2021analyzing, freire2019characterization}.
By applying natural language processing tools, researchers have studied the general discussion around the migration topic in Twitter leveraging, e.g., the expressed sentiment~\cite{ozturk2018sentiment} or patterns in the language that indicate empathy or a perceived threat~\cite{freire2021framework}.

When we focus on online sociopolitical debates, analyzing polarizing debates requires automatic classification of users' stance towards a given entity, like a controversial topic (e.g., migration).
Twitter has been a source for multiple stance annotated datasets covering political topics and social aspects.
One of the most popular was build for the SemEval-2016-Task 6~\cite{DBLP:conf/semeval/MohammadKSZC16} and includes 4,163 tweets around 5 topics annotated with stance and sentiment.
\citeauthor{ghosh2019stance} explored the reproducibility of existing stance detection models on this dataset~\cite{ghosh2019stance} and showed that a pre-trained BERT~\cite{devlin2018bert} model can outperform many existing methods on this task.
Another dataset contains 9,973 tweets related to the ``MeToo'' movement, annotated for five different linguistic aspects~\cite{gautam2020metooma}.

While the majority of existing datasets and stance analyses focused solely on English content, a few cover a different language, and even less extent to multiple languages.
On the issue of Catalan independence, two related multilingual datasets of Spanish and Catalan tweets were created~\cite{DBLP:conf/sepln/TauleMRRBP17, DBLP:conf/sepln/TaulePMR18}.
The ``SardiStance'' dataset contains 3,242 Italian tweets related to the \emph{Sardines movement}~\cite{DBLP:conf/evalita/CignarellaLBPR20}.
Two datasets on the 2016 referendum on the reform of the Italian Constitution have been created~\cite{lai2018stance, lai2020multilingual}.
Additionally, tweets concerning the French 2017 presidential elections~\cite{lai2020multilingual} were annotated for stance towards the candidates.
\citeauthor{graells2020every} studied the debate of abortion in Chile and Argentina between 2015 and 2018~\cite{graells2020every}.
The authors rely on keywords in profile information and tweets to infer a user's stance and build a classifier to propagate it to the rest of the dataset.
\citeauthor{darwish2018predicting} labeled the attitude of users towards Muslims after the Paris terrorist attacks in 2015 and analyzed the applicability of historical tweets and network features to detect this stance pre-event, assuming it stays constant~\cite{darwish2018predicting}.
However, while the post-event analysis included multiple languages, only English tweets were considered for the pre-event detection.

In this paper, we do not treat each tweet as a standalone document, but use Twitter conversations where a news tweet sets the topic. 
Exploiting the structure given by Twitter conversations was shown to benefit rumour stance classification~\cite{zubiaga2016stance}.
Further, \citeauthor{villa2020stance} %
noted a difference in the handling of the modalities of replies and quotes~\cite{villa2020stance}.
\citeauthor{kumar2021weakly} used Hashtags to infer a user's stance and propagate it as weak labels to Twitter replies~\cite{kumar2021weakly}.
Stance detection on Twitter so far mainly focuses on individual posts or network features, or relies on weak supervision through keywords and hashtags.
In our work, we annotate the stance towards migration in replies to related news on Twitter in five languages.
We use news as context without relying on specific language or hashtags in the replies.
We further use the annotations to interpolate the public's stance in Europe over one year.

\section{Twitter News Discussion Dataset}
\label{Sec:DataSet}
In this section, we describe the collection of migration related news tweets with replies and the subsequent stance annotation.

\subsection{Content Collection}
\label{subsec:content-collection}
Discussions on migration can be diverse, e.g., referring to national and international events or ongoing policy debates.
Additionally, users can express an attitude on social media toward migration without using specific keywords.
This makes it impractical to filter posts only by hashtags or keywords.
Instead, we rely on the broad topic coverage of news media outlets.
We gathered an extensive list of media outlets in European countries with an active presence on Twitter.
In particular, we focus on the five western and central European countries France, Germany, Italy, Poland, and Spain.
These countries are affected to varying extents by the crisis in Ukraine due to diverse closeness and the number of refugees.
Poland has also recently been the subject of two large refugee crises. %

For news outlets accounts, we rely on information from the open knowledge base Wikidata\footnote{\url{https://www.wikidata.org/}}.
This readily available source stores claims and facts about entities in a structured format.
The employed list is based on a Wikidata database dump from March 2022. %

First, the complete dump is filtered by entities containing a Twitter ID property.
For narrowing down to account identifiers of media outlets, we use the \texttt{instance\_of} property that categorizes each entity.
We look for entries where this property matches values such as ``newspaper'', ``state media'', or ``television channel''.
Applying those filters, we collected  -- worldwide -- a total of 8,159 media-related entities, each with at least one associated Twitter account.
A subsequent examination of country-specific results showed different degrees of coverage, requiring additional effort to represent the media landscape of a country as comprehensively as possible.
Starting from the individual accounts in each country, we used Twitter's recommendation feature to manually discover additional identifiers of media sources for the European countries.
Combined, this results in 565 news outlet accounts for this paper's five countries of interest (i.e., France, Germany, Italy, Poland, Spain).

We rely on the Twitter API to collect all posts made by these accounts in one year, starting on 1st September 2021. 
This totals over 5.5M tweets (referred to as ``all'' in Fig.~\ref{fig:news-sentiment}).
Due to limits imposed by the API, we only collect replies for posts with $\geq$ 5 replies.
To help with filtering and the subsequent annotation process, we make use of the NLLB-200~\cite{nllb} translation model to translate each tweet into English. %
We select news posts by the following keywords to extract only migration-related news: \emph{refugees, refugee, migrant, migrants, migration, immigrant, immigrants, immigration}.
Given the rather formal and relatively neutral language used by the news media (even on social media), we assume that these keywords are coherent with most of the coverage of the issue.
Lastly, only replies written in the local language of each country are selected in order to limit ourselves as much as possible to the responses of the respective populations.
Such filters would be inefficient for English-speaking countries (e.g., UK).
The final dataset (see Tab.~\ref{table:twitter-dataset}) contains 11,478 posts from 200 news outlets together with 146,711 replies.

\begin{table}[t]
    \caption{Twitter dataset overview. Migration-related tweets from news outlets and corresponding replies per country.}
    \small
    \centering
    \setlength{\tabcolsep}{4pt}
    \begin{tabular}{lcccc}
        \toprule
        \textbf{Country}&\textbf{News Outlets}&\textbf{News Tweets}&\textbf{Replies}&\textbf{Users}\\
        \midrule
France & 37 & 2,020 & 32,839 & 17,433 \\
Germany & 72 & 3,752 & 55,317 & 20,461 \\
Italy & 21 & 1,305 & 9,892 & 5,837 \\
Poland & 35 & 3,138 & 27,892 & 12,578 \\
Spain & 35 & 1,263 & 20,771 & 13,823 \\
\midrule
& 200 & 11,478 & 146,711 & 70,132\\
        \bottomrule
    \end{tabular}
    \label{table:twitter-dataset}
\end{table}

\subsection{Labeling of Stance Training Data}\label{sec:stance-annotation}
The manual annotations process of training data for preparing stance classification was handled internally by two researchers to ensure a high quality of annotations.
We specifically decided against crowd-sourcing due to the data's sensitivity and the labeling complexity.
Each news and reply pair was presented with the original text, a translation, and a URL to the tweet. %
At first, both researchers studied and discussed a sample of the data and refined the annotation ruleset.
For a given reply tweet, the expressed stance toward migrants and refugees can be categorized into three classes: \emph{Positive}, \emph{Negative}, \emph{Neutral}.
\emph{Positive} typically included tweets that showed empathy for the displaced persons, willingness to help, or appreciation for supportive politics.
In the \emph{Negative} case, there is a demand to protect the border, fear of exploitation of the social system, or a call for deportation.
This includes tweets that are dismissive of only one group of refugees while accepting the other (e.g., second negative tweet in Tab.~\ref{table:example-tweets}).
In both cases, the news tweet sets the context as shown in Tab.~\ref{table:example-tweets}. 
Even replies with a negative tone or insulting content can be considered \emph{Positive} if statements hostile to refugees are being opposed.
The default class is \emph{Neutral}.

To obtain a broad coverage, we selected 500 random news tweets, including one reply for each of the countries France, Germany, Italy, and Spain from the dataset shown in Table~\ref{table:twitter-dataset}.
We slightly modified the selection process for Poland, as it was affected by two major migration movements during the period under consideration. 
First, in the second half of 2021, by the Belarus-European Union border crisis and, from the end of February 2022, by the refugee flow from Ukraine.
To take a closer look at the impact of these two events, we focused on two time periods (roughly one month each) linked to those events.
For those periods, we selected all Polish news tweets and corresponding replies for manual labeling. 
For the Belarus crisis, this is the period from Nov. $7^{th}$ to Dec. $7^{th}$, 2021, and for the Ukrainian refugees starting from the beginning of the war on Feb. $24^{th}$, 2022 to the end of March.
We refer to these datasets as \emph{Poland BY} and \emph{Poland UA}, respectively.

To ensure consistent annotations, parts of the Polish datasets were processed by both researchers.
We specifically used these subsets in the inter-agreement verification because we anticipated a media reframing and an audience stance shift between the periods they represent.
We evaluated inter-annotator agreements using Krippendorf's alpha (K-$\alpha$).
The \emph{Poland UA} dataset contains $506$ common annotations with a strong agreement of $0.801$.
For \emph{Poland BY} we achieve an agreement of $0.805$ with $350$ annotations.
Following~\cite{artstein-poesio-2008-survey}, we assume good annotation quality.
Once consistency was established, one researcher at a time performed the remaining annotations of each dataset.
Further, it is worth mentioning that the German and Spanish datasets were processed by native speakers (the two researchers), while other languages mainly required English translation as outlined in Sec.~\ref{subsec:content-collection}.
Table~\ref{table:annotations} provides an overview of the annotated datasets.

\begin{table}[t]
    \caption{Examples of different stance towards migrants or refugees in \noindent {\color{purple} replies} to \noindent {\color{teal} news tweets} (translated).}
    \footnotesize
    \centering
    \setlength{\tabcolsep}{4pt}
    \begin{tabular}{p{39.5mm}p{39.5mm}}
        \toprule
            \textbf{Positive Tweet Reply Stance} & \textbf{Negative Tweet Reply Stance} \\
        \midrule
        \noindent {\color{teal} According to several aid agencies, many refugees could soon leave Ukraine for the West. Germany is also preparing.} [SEP]
        \noindent {\color{purple} I hope we don’t get to feel the same welcome culture as 2015. That was extremely embarrassing.} & 
        \noindent {\color{teal} According to the report, the federal government wants to bring hundreds of refugees from Afghanistan to Germany. We could get 200 people to safety every week.} [SEP] 
        \noindent {\color{purple} Go to the chancellery in Berlin, set up tents.} \\
        \midrule
        \noindent {\color{teal} “This is our moral duty”: President Ursula explains the temporary protection mechanism to provide quick protection to Ukrainian refugees. European border management facilitates formalities at the borders.} [SEP] 
        \noindent {\color{purple} What a double standard.} & \noindent {\color{teal} The first refugees from Ukraine have now arrived in Germany, for example in Freiburg, where more than 150 children from the vicinity of Kiev arrived.} [SEP]
        \noindent {\color{purple} None of these real refugees will want to kill you in the name of their god.} \\
        \bottomrule
    \end{tabular}
    \label{table:example-tweets}
\end{table}

\begin{table}[t]
    \caption{Annotated datasets of replies to news tweets regarding stance towards migrants or refugees.}
    \small
    \centering
    \setlength{\tabcolsep}{4pt}
    \begin{tabular}{lccccc}
        \toprule
        \textbf{Dataset}&\textbf{Annotator}&\textbf{Replies}&\multicolumn{3}{c}{\textbf{Stance}}\\
        & & &\textbf{Positive}&\textbf{Negative}&\textbf{Neutral}\\
        \midrule
Poland UA & A & 3,368 & 1,628 & 306 & 1,434 \\
Poland BY & A & 2,874 & 345 & 1,375 & 1,154 \\
France & A & 500 & 131 & 240 & 129 \\
Germany & A & 500 & 94 & 305 & 101 \\
Italy & B & 500 & 136 & 226 & 138 \\
Spain & B & 500 & 141 & 184 & 175 \\
        \midrule
 &  & 8,242 & 2,475 & 2,636 & 3,131 \\
        \bottomrule
    \end{tabular}
    \label{table:annotations}
\end{table}

\subsection{Dataset Sharing}
\label{sec:data_sharing}
Following the FAIR data principles, we publish this dataset on Zenodo\footnote{\anon[Zenodo URL omitted for review]{\url{https://doi.org/10.5281/zenodo.7189338}}}.
To comply with Twitter's terms and conditions, we do not provide the full tweet JSON but share tweet IDs that can be rehydrated.
The dataset reflects the data shown in Table~\ref{table:twitter-dataset} and \ref{table:annotations}.
It consists of news outlet usernames, tweet/reply pairs, and the reply IDs used in the annotation process.
Due to their sensitive nature, we only disclose the associated stance labels upon request.

\section{Methodology}\label{Sec:Meth}

Our approach to studying a potential change in European attitude toward migration after the beginning of the war in Ukraine is twofold: we analyze the reflection in mass media and the effect of these events on the public's attitude.
We use data collected from social media discussions for this purpose.
To measure the shift in the public debate, we make a contrastive analysis using content from before and after February 2022.
We base our research on the language processing of tweet content through sentiment detection, stance detection, and the terminology used.

The novelty of our approach to stance detection lies in its systematic multilingualism and in the context-dependent way the content is analyzed. %
To measure a user's targeted attitude toward migrants and refugees, we do not rely on keywords or hashtags to select suitable responses, but instead leverage the topics set by the news.

\subsection{Sentiment Analysis}
As a first measure, we use sentiment detection to examine differences in the nature of reporting and subsequently the relationship to public stance.
To this end, we annotate each tweet using the sentiment analysis tool VADER~\cite{hutto2014vader}.
It is tuned for sentiment in microblog-like social media and employs lexical features together with grammatical and syntactical features for measuring sentiment intensity.
While it only supports English, previous studies~\cite{balahur2012multilingual} have shown that the performance penalty on translated text for sentiment analysis is acceptable.
Applied to translated tweets, it provides a normalized, weighted composite score in the range of $-1$ to $1$.
Negative values correspond to a negative sentiment and vice versa.

\begin{figure}[t]
    \centering
    \includegraphics[width=\columnwidth]{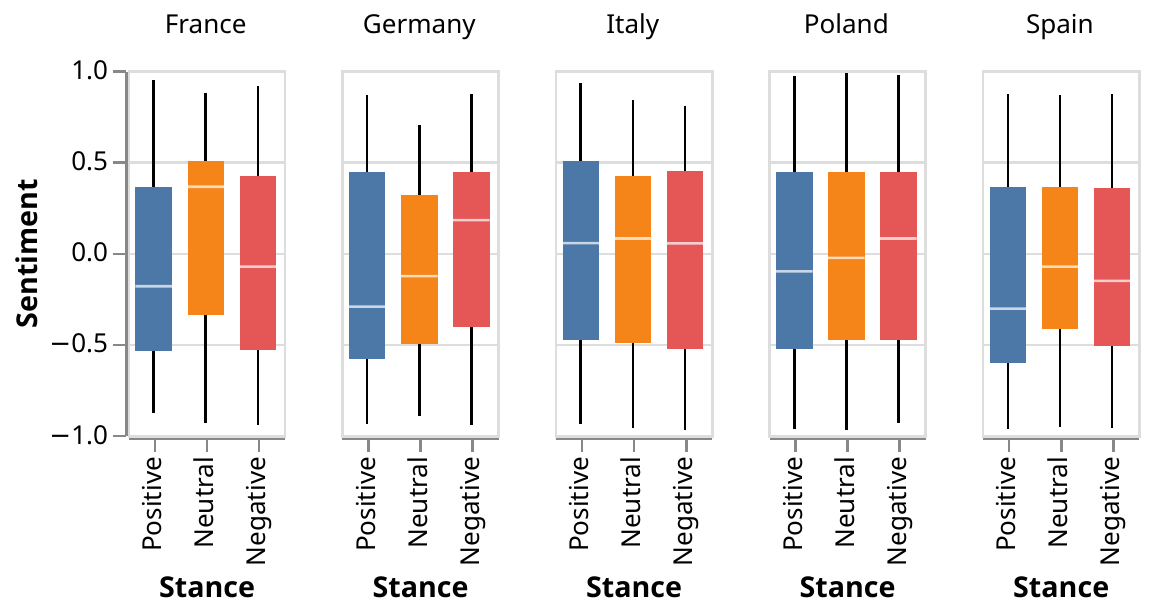}
    \caption{Reply sentiment distribution per stance label by countries in annotated datasets.}
    \Description{Sentiment of replies does not correlate to the stance they show toward migrants. Positive stance on average even has a slightly negative sentiment.}
    \label{fig:sentiment-stance-replies}
\end{figure}

Examining the sentiment distribution for the stance annotated replies in the datasets (see Fig.~\ref{fig:sentiment-stance-replies}), no clear relationship between sentiment and stance labels emerges.
This is in line with the conclusions of previous works~\cite{mohammad2017stance} that sentiment alone is not an indicator of stance.
However, in the following analyses, we are interested in the effects of sentiment in news on the audience's stance.

\subsection{Language Analysis}
\label{sec:rank_diff}
We set to empirically investigate the language shift in news media coverage towards migration after the beginning of the Russian invasion of Ukraine.
This addresses the questions if and how news outlets reframe their discourse on this topic.
To tackle this task, we use the {\em rank difference} method proposed in~\cite{jbp:/content/journals/10.1075/term.14.2.05kit}.
The rank difference algorithm is used to identify terms that characterize a specific text corpus against a background reference collection.
For example, the word \textit{court} will probably be recognized as a characteristic term if we analyze a corpus of legal documents.
The method creates two ranked lists of terms based on their frequency in the \emph{foreground} domain and the \emph{background} corpus.
Afterwards, it identifies terms with the largest difference by comparing their relative position in both rankings.
In other words, they are used considerably more often in the analyzed domain.
These unusual word frequencies will indicate the importance of these terms in the foreground domain.
The formula for calculating rank difference is shown in Equation \ref{eq:rank_diff}, 

\begin{equation} \label{eq:rank_diff}
\tau(w) = \frac{r_D(w)}{\sum_{w' \in V_D}r_D(w')} - \frac{r_B(w)}{\sum_{w' \in V_B}r_B(w')}
\end{equation}

\noindent where $r_D(w)$ and $r_B(w)$ are the ranks of term $w$ in the foreground and background corpus, respectively.
To account for the potential difference in vocabulary size between domains, a rank normalization is applied against the summation of all term rankings in the corresponding vocabulary ($V_D$ and $V_B$).
Then, we understand $\tau(w)$ as the relevance score for the term $w$ in the foreground compared with the background corpus. This method has been shown to help characterize a community stance in online discussions of polarizing topics, both in a general context and for news media~\cite{10.1145/3201064.3201076, polquiz_politicalbias}.

\subsection{Stance Classification}\label{sec:method-classifier}
Transformer-based models such as BERT~\cite{devlin2018bert} are established as state-of-the-art for a wide range of classification tasks.
Available models pre-trained on large corpora of unsupervised texts achieve top performance by further tuning on task-specific datasets.
For multilingual use cases, adapted models such as M-Bert or XLM-RoBERTa~\cite{xlm-roberta} have been developed.
In the domain of stance recognition, BERT can outperform both feature-based and other deep learning approaches in a monolingual setting~\cite{ghosh2019stance}.
Further, \citeauthor{xstance} used M-BERT to investigate the applicability of cross-lingual and cross-target transfer of stance detection~\cite{xstance}.

As our datasets originate from Twitter and are multilingual, we select the XLM-RoBERTa model \emph{XLM-T}~\cite{xlm-t} that is pre-trained on around 198M multilingual tweets\footnote{\url{https://huggingface.co/cardiffnlp/twitter-xlm-roberta-base}}.
Within the context of BERT, we interpret the classification of tweet-reply pairs as sequence pair classification~\cite{kumar2021weakly}.
Tweet and reply are separated with the special token \texttt{[SEP]} and prefixed with the token \texttt{[CLS]}.
A linear layer is added on top to transform the final hidden state of \texttt{[CLS]} into a probability for the stance classes.
For tuning the model on the individual datasets, we use a batch size of 16, a maximum sequence length of 128, and perform a Bayesian search over the following hyperparameters to maximize the macro F1 score:
\begin{itemize}
    \item Number of epochs: 2, 3, 4, 5
    \item Learning rate: 2e-5, 3e-5, 4e-5, 5e-5
    \item Learning rate for class. layer: 1e-3, 2e-3, 3e-3, 4e-3, 5e-3
\end{itemize}
For the remaining parameters we follow the recommendations for fine-tuning BERT: warm-up proportion 0.1, linear decay of the learning rate, Adam with $\beta$1 = 0.9 and $\beta$2 = 0.999, and weight decay of 0.01 for all layers except 'bias'.
Each parameter selection is tested via 5-fold cross-validation.
Stratified sampling is applied to ensure folds preserve the percentage of samples for each class.

We use the tweet text in the original language without translation.
Preprocessing consists of removing URLs and mentions (@) plus the removal of the hashtag prefix (\#).
We decided against the removal of the complete hashtags, because in this case they are often part of a sentence (e.g., ``After all, they are not \#refugees but \#tourists.'') and only occur in 157 replies of the annotated dataset.

\section{Results and Discussion}
\label{Sec:Results}
In this section, we first analyze the media coverage during the selected period by looking into variations of the average sentiment over time and the language used.
Then, we present our findings regarding public opinion through our proposed stance detection models.
Finally, we look into potential interactions between media coverage and public opinion as expressed in social media.

\subsection{Media Analysis}
\label{subsec:media-analysis}
\begin{figure}[t]
    \centering
    \includegraphics[width=\columnwidth]{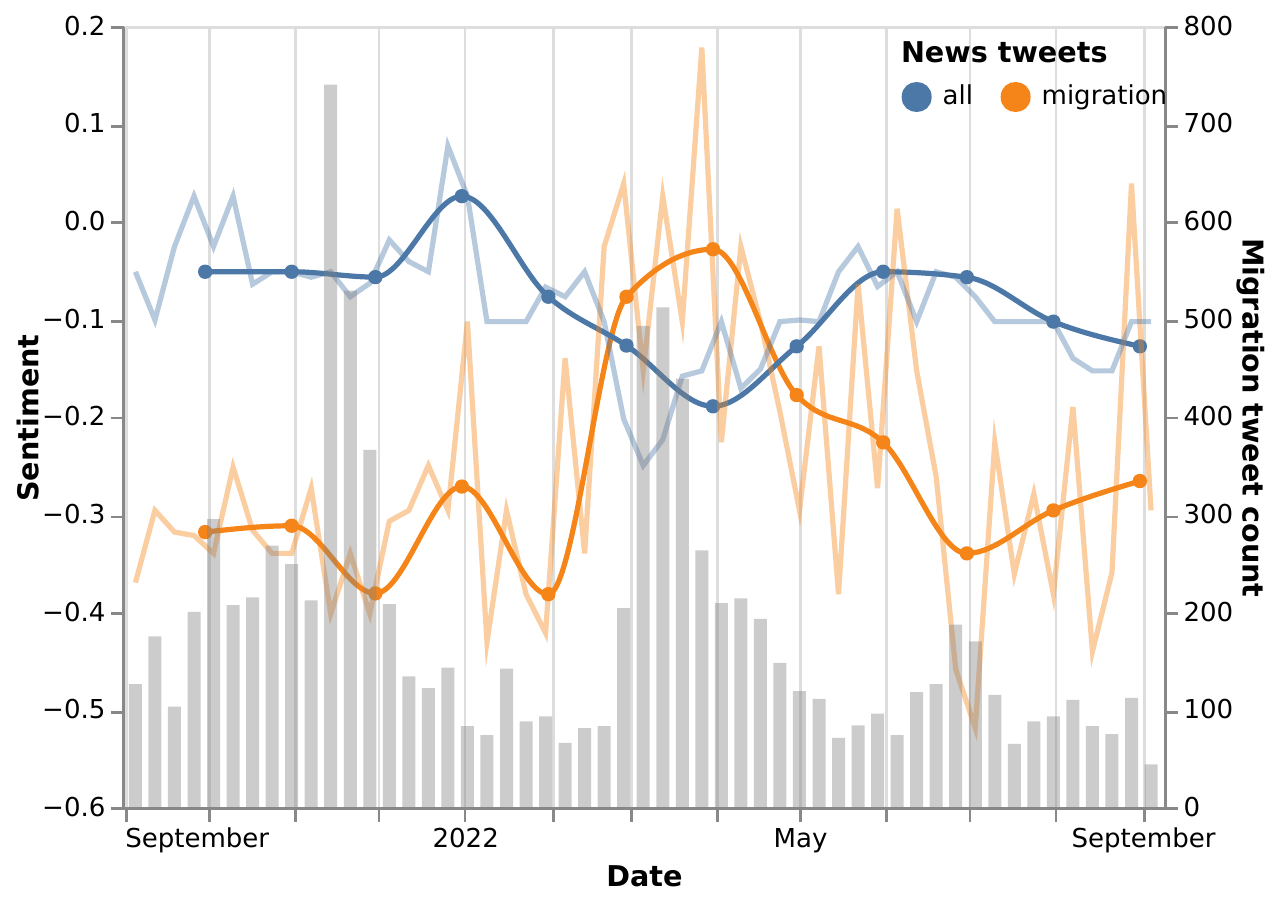}
    \caption{Median sentiment of all (blue) and migration related news tweets (orange) \& migration related news volume (bars).}
    \Description{The sentiment of news tweets on migration is generally more negative than the average. However, in the first two months after the Ukraine invasion, migration related news are significantly more positive (from around -.3 to 0), even surpassing the general reporting.}
    \label{fig:news-sentiment}
\end{figure}

In Figure~\ref{fig:news-sentiment}, we contrast the aggregated European news outlets' sentiment on migration-related topics versus all topics during the one-year period under review.
The median is formed monthly (smooth lines) as well as weekly without including neutral tweets (i.e., sentiment=0).
In general, coverage of migration issues is significantly more negative than average.
However, the effects of the Russian invasion of Ukraine in February $24^{th}$, 2022 and the subsequent refugee movement are clearly visible in both curves.
As a result, the general news coverage seems to be overtaken by the adverse reports of the war, while the refugee subject is conveyed much more positively.
This is particularly evident in Poland, where news sentiment about migration swings sharply to the positive (see Fig.~\ref{fig:news-sentiment-reply-stance-countries} in the Appendix).
In contrast, the average for the five analyzed countries (see Fig.~\ref{fig:news-sentiment}) goes only to neutral values.

Regarding the volume of migration-related news, we see in Figure~\ref{fig:news-sentiment} an increase (>100 news/week) for the periods in which European countries were faced with larger migration-related incidents.
For example, the last months of 2021 were marked by the Belarus-EU border crisis, with a peak in November.
Another significant increase follows after the start of the Russia-Ukraine war.
We, therefore, assume that filtering news by keywords as described in Sec.~\ref{subsec:content-collection} captures enough evidence to reflect the discussion on migration.

\subsection{Nature of the Media Coverage Shift}
The sentiment analysis on the news coverage already shows a shift in the media's attitude when covering the topic.
In this section, we investigate the nature of this shift in terms of the vocabulary used.

As the corpus for our foreground domain, we used tweets from news outlets expanding one month after the beginning of the Russia-Ukraine war (from March $1^{st}$ until March $31^{st}$, 2022).
For the background corpus, we use one month of news outlets' tweets from November 2021 (from Nov. $1^{st}$ until Nov. $30^{th}$, 2021).

We use bi-grams as terms and calculate the rank difference for each language.
We prefer bi-grams over words because they offer more context, making it easier for humans to assess the connotation of terms within the discussion.
We preprocessed the English translation of the tweets' text and remove punctuation marks, stopwords, user mentions and URLs.
Also, we decided to remove named entities (e.g., places, politicians, news outlets) because, although they might indirectly hint at a stance, these are very circumstantial and do not directly contribute to characterizing a general shift in the language used.
Finally, we lemmatized the remaining words.

Table~\ref{table:rank-diff} shows the top 10 most characteristic terms for each country and period.
We see a general trend among the selected media to reframe the migration-related discussion from ``migrant'' to ``refugee''.
This difference is sometimes stressed with phrases like ``real refugee'' or ``war refugee'', bringing up again the question of legitimacy in the protection claim from different groups \cite{doi:10.1080/1369183X.2017.1348224, long2013refugees}.
In general, the concerns in the media discourse before the war seemed to be dominated by ethnocentrism or concerns around the effect of migration on society (this is especially visible in Spain with terms like ``great replacement'' or ``noneuropean population'').

The desecuritization of the discourse is also evident from the list of terms.
In 2021, multiple references point to issues like police incidents, state of emergency, or attacks.
On the other hand, the 2022 sample concentrates more on transmitting a welcoming message and the need for aid/help.
Even the security concerns in this second period appear to be focused on the protection of the refugees (e.g., ``attacked refugee'', ``refugee child'').
Another relevant aspect across various countries is the racism/xenophobia dimension (exemplify with terms like ``racist stereotype'', ``white christian'', or ``racist xenophobic'').
Interestingly, for some countries like France and Spain, this discussion seems triggered by the new situation in Europe, while in Germany, this was prominent before the war.

\begin{table*}[ht]
    \caption{Rank-Difference of media language before and after the Ukrainian crisis.} 
    \footnotesize
    \centering
    \begin{tabular}{cccccc}
        \toprule
        \textbf{Period}&\textbf{Poland}&\textbf{France}&\textbf{Germany}&\textbf{Italy}&\textbf{Spain}\\
        \midrule
    \multirow{10}{*}{November 2021} &emergency zone &rebuild peace &migration background &vision wall &great replacement \\
    &state emergency &friendship treaty &drowned atlantic &brother migration &native population \\
    &detained police &africa improve &migrant drowned &migration inevitable &population declining \\
    &tried escape &improve integration &dozen migrant &migrant sea &population arriving \\
    &border detained &want rebuild &patient migration &eu-funded wall &come dissolve \\
    &police unknowingly &migrant want &border wall &wall violated &growing native \\
    &filming documentary &integration migrant &route refugee &stop migrant &continent growing \\
    &unknowingly entered &project immigration &quota attack &government right &conspiracy theory \\
    &entered state &shortage biomedicine &racist stereotype &want throw &noneuropean population \\
    &zone tried &avoid shortage &attack youth &rule migrant &victim migration \\
    \midrule
    \multirow{9}{*}{March 2022} &million refugee &mea culpa &refugee war &real refugee &dead man \\
    &economic situation &proimmigrant actor &minister health &refugee go &war refugee \\
    &train station &comment migrant &plan offer &refugee apply &invasion young \\
    &aid worker &writer proimmigrant &vaccination refugee &war real &refugee welcomed \\
    &refugee center &lost writer &refugee child &refugee fleeing &difference invasion \\
    &worker outside &attack lost &willingness help &help refugee &understand difference \\
    &station reporter &attack comment &refugee regardless &voted yes &orderly immigration \\
    &attacked refugee &migrant pathetic &register refugee &refugee welcomed &migrant attempt \\
    &men attacked &white christian &uncontrolled entry &send weapon &racist xenophobic \\
        \bottomrule
    \end{tabular}
    \label{table:rank-diff}
\end{table*}

\subsection{Stance Detection and Cross-linguality}
\label{subsec:stance-detection-cross-linguality}
\begin{table}[ht]
    \caption{Performance of stance classifiers on individual datasets. Scores are cross-validated accuracy (Acc) and macro F1. Baseline is given by a ZeroR classifier.}
    \small
    \centering
    \setlength{\tabcolsep}{4pt}
    \begin{tabular}{l|cc|cc}
        \toprule
        &\multicolumn{2}{c}{\textbf{XLM-T}}&\multicolumn{2}{c}{\textbf{ZeroR}}\\
        \textbf{Dataset}&\textbf{Acc}&\textbf{F1}&\textbf{Acc}&\textbf{F1}\\
        \midrule
Poland UA & 0.70 & 0.62 & 0.48 & 0.22 \\
Poland BY & 0.62 & 0.52 & 0.48 & 0.22 \\
Poland UA/BY  & 0.67 & 0.67 & 0.41 & 0.20 \\
FR / DE / IT / ES  & 0.54 & 0.48 & 0.48 & 0.22 \\
France & 0.54 & 0.50 & 0.48 & 0.22 \\
Germany & 0.63 & 0.49 & 0.61 & 0.25\\
Italy & 0.54 & 0.52 & 0.45 & 0.21 \\
Spain & 0.50 & 0.49 & 0.37 & 0.18 \\
        \bottomrule
    \end{tabular}
    \label{table:classification-performance}
\end{table}

\begin{table}[ht]
    \caption{Cross-lingual stance classification performance. Model tested on languages not present in the training set.}
    \small
    \centering
    \setlength{\tabcolsep}{4pt}
    \begin{tabular}{ll|cc|cc}
        \toprule
        &&\multicolumn{2}{c}{\textbf{XLM-T}}&\multicolumn{2}{c}{\textbf{ZeroR}}\\
        \textbf{Training Dataset}&\textbf{Test Dataset}&\textbf{Acc}&\textbf{F1}&\textbf{Acc}&\textbf{F1}\\
        \midrule
UA BY DE IT ES & FR & 0.51 & 0.49 & 0.48 & 0.22 \\
UA BY FR IT ES & DE & 0.55 & 0.49 & 0.61 & 0.25\\
UA BY FR DE ES & IT & 0.53 & 0.51 & 0.45 & 0.21 \\
UA BY FR DE IT & ES & 0.50 & 0.50 & 0.37 & 0.18 \\
FR DE IT ES & UA BY & 0.42 & 0.41 & 0.41 & 0.20 \\
UA BY & FR DE IT ES & 0.46 & 0.40 & 0.48 & 0.22 \\
        \bottomrule
    \end{tabular}
    \label{table:cross-classification-performance}
\end{table}

To investigate the public's stance toward migrants and refugees, we employ an adapted classifier on all available news replies.
Using the annotated datasets (Sec.~\ref{sec:stance-annotation}), we first acquire an understanding of the accuracy that can be expected and investigate the generalizability to other languages.

To this end, we train the model described in Sec.~\ref{sec:method-classifier} on different combinations of country-specific datasets.
Table~\ref{table:classification-performance} provides an overview of the 5-fold cross validated classifier performance.
The corresponding confusion matrices are supplied in Appendix~\ref{appendix:confusion}.
As a baseline, we include a Zero Rate (ZeroR) classifier that always predicts the majority class.
Since our datasets for the other countries have a smaller coverage compared to that of Poland, we additionally use the option of pre-training a model with the Polish data in these cases and fine-tuning with the target country.
For France, Germany, Italy, and Spain, we reach a similar performance of around $0.5$ F1.
Combining the four datasets also seems to provide no additional benefits in this context.
However, these scores align with results achieved in other works for stance detection in Twitter conversations~\cite{villa2020stance, kumar2021weakly, zubiaga2016stance}.
While the value for \emph{Poland BY} is also in the same range, for \emph{Poland UA}, we achieve a significantly higher F1 of $0.62$.
Additionally, we see a clear advantage of combining both datasets, for which a score of $0.67$ is subsequently achieved.

Furthermore, we look at the transferability to other languages.
We train a model on all languages except the one used for testing.
The results of this cross-lingual stance classification are presented in Table~\ref{table:cross-classification-performance}.
Notably, the tests on the individual languages without Polish achieve a similar result as the previous tuning on a single dataset.
In addition, we see a significantly lower performance of about $0.4$ F1 in the tests when using the Polish datasets on one side (Training/Test) and the remaining countries on the other.
This suggests a greater commonality between the topics in the datasets of France, Germany, Italy, and Spain.
Ultimately, these experiments demonstrate that a model tuned for this task can be sufficiently generalized to be applied to additional languages or countries.

\subsection{Public Stance Towards Migration}
\begin{figure*}[ht]
    \centering
    \includegraphics[width=\textwidth]{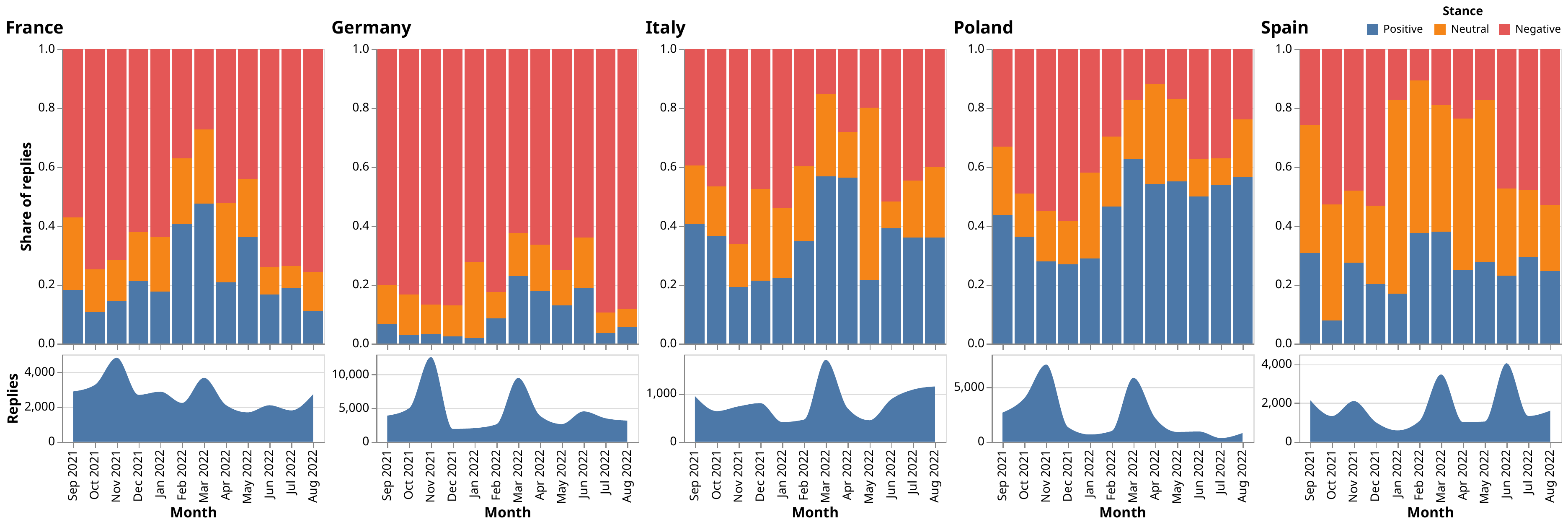}
    \caption{Breakdown of stance toward migrants and refugees in replies to news tweets per country and number of replies.}
    \Description{In each of the 5 countries, there is an increase in the proportion of positive attitudes accompanied by a decrease in negative attitudes in replies starting February 2022.}
    \label{fig:stance-countries}
\end{figure*}
This section will examine the responses to the relevant news reports on social media.
We are analyzing the public stance toward migration in individual countries. 
For this purpose, as outlined in the previous section, we use a classifier pre-trained on the Polish datasets and fine-tuned with the combined data from the four other countries.
This approach is intended to prevent a possible bias towards Poland's significantly larger proportion of data.
We then use the trained model on all news replies across the dataset presented in Table~\ref{table:twitter-dataset}.
Figure~\ref{fig:stance-countries} shows the monthly aggregated relative frequency of stance labels per country.
In addition, we show -- in the lower part of the figure -- the volume of replies in each month, reflecting public attention to the migration issue.

Comparing the distribution of the general attitude, Poland is the most positive while the tendency of Germany is the most negative. 
However, since the positive class is underrepresented in our labeled data for the countries excluding Poland, we cannot rule out a bias in the classifier towards a negative stance.
So, we focus more on the relative changes.
Moreover, this analysis targets only social media and is therefore subject to selection bias, meaning it cannot be applied unreservedly to the views of the general population.

We can observe a noticeable stance shift in the positive direction for all countries at the beginning of the Ukraine war, starting in February 2022.
This can be seen most clearly in Poland, which was also the country most affected by the subsequent refugee movement.
There, a lasting change in attitude is also evident, while the effect in the other countries is less stable and fades after about 3 to 4 months. 
At the same time, the implications of the Belarus crisis in Poland are also visible, accompanied by a noticeably negative stance.

The peaks in the reply volume correlate with the news volume peaks shown in Figure~\ref{fig:news-sentiment}.
This is to be expected, as higher news density leads to more comments.
Both can be considered as signals of public attention to the respective topic, whereas social media gives a stronger grassroots signal.
For the culmination of the Belarus crisis in November 2021 and the Ukraine war starting in February 2022, increased attention can be observed in all countries.
Spain marks another high in June due to the death of a migrant at the border between the Spanish enclave of Melilla in Morocco and the subsequent protests against European migration policies.

This analysis reveals that the trained model and approach used for stance detection, can reflect tendencies in public opinion and identify incisive events.

To further investigate the interactions between mass media and the public's attitude, we perform a Granger causality test~\cite{granger1969investigating}.
We use the weekly average media sentiment and the average reply stance time series (see Figure \ref{fig:teaser}).
We found that the media sentiment leads the public stance ($F=26.11, p<.0001$), with the one-week lag showing the highest F-statistic.
This might indicate a potential influence of media coverage on the audience's attitude.
However, although weaker, we also found that replies stance granger-cause the sentiment in the media coverage ($F=12.74, p<.001$).
The results may suggest a feedback loop effect on the media based on public perception~\cite{soroka2015s}.

\section{Conclusions}
\label{Sec:Conclusions}
Coming back to the original question from the title ``Migration reframed?'', we have shown that there actually was a measurable shift in sentiment in media coverage on migration combined with different wording.
The same is reflected in the stance of the audience.
This, however, was only of limited duration for many countries.
So, in summary, we can say the discussion was reframed, including a change in the language used in the context of migration, but in a rather temporal, event-driven fashion. 

From the technology side, our multilingual and contextual stance detection method has proven very effective for the study performed.
It eases the comparison of situations across multiple countries, which can yield further insights into societal processes such as migration.
Moreover, it suggests a potential generalizability to other languages.
Using our labeled dataset, we showed that sentiment is not an appropriate indicator to measure audience's opinion on such topic.
Furthermore, our indirect data collection method based on news tweets has allowed us to capture social media discussions on a topic without depending on keywords in the comment itself. 

Our work offers various directions for future sociology and technology research.
Examples include a deeper analysis of the reasons for the differences between European countries, distinction in the interaction of individual media with their audience, and narratives around migration.

\begin{acks}
This paper is part of a project that has received funding from the \grantsponsor{EU}{European Union}{https://doi.org/10.3030/101021866}'s Horizon 2020 research and innovation programme under grant agreement No. \grantnum[]{EU}{101021866}. %
\end{acks}

\bibliographystyle{ACM-Reference-Format}
\bibliography{references}

\appendix
\newpage
\section{News Sentiment and Reply Stance}
\label{appendix:sentiment_vs_stance}
\begin{figure}[h]
    \centering
    \includegraphics[width=\columnwidth]{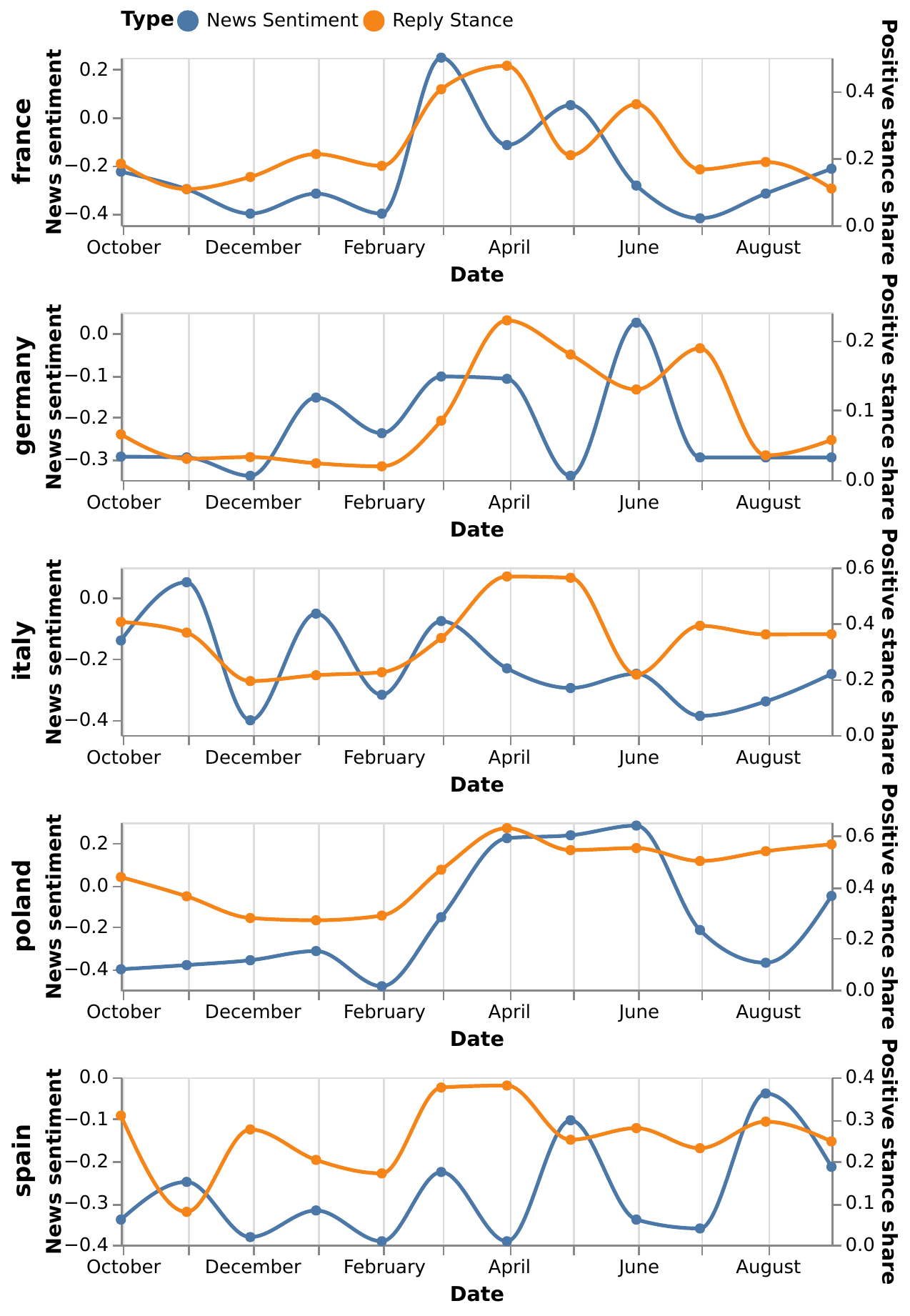}
    \caption{Monthly median news sentiment and positive reply stance share timelines by country.}
    \label{fig:news-sentiment-reply-stance-countries}
    \Description{Each country shows a significant uptick in positive reply stance after February 2022. For news sentiment, however, this is not uniform across countries.}
\end{figure}

As an extension to the media sentiment analysis performed in Section~\ref{subsec:media-analysis} and the aggregated data for Europe shown in Figure~\ref{fig:news-sentiment} we provide a faceted view on this matter divided by country in Figure~\ref{fig:news-sentiment-reply-stance-countries}.
For better comparability, the respective reply stance timelines are included. 

\newpage
\section{Stance Classification Confusion}
\label{appendix:confusion}
\begin{figure}[h]
    \centering
    \includegraphics[width=\columnwidth]{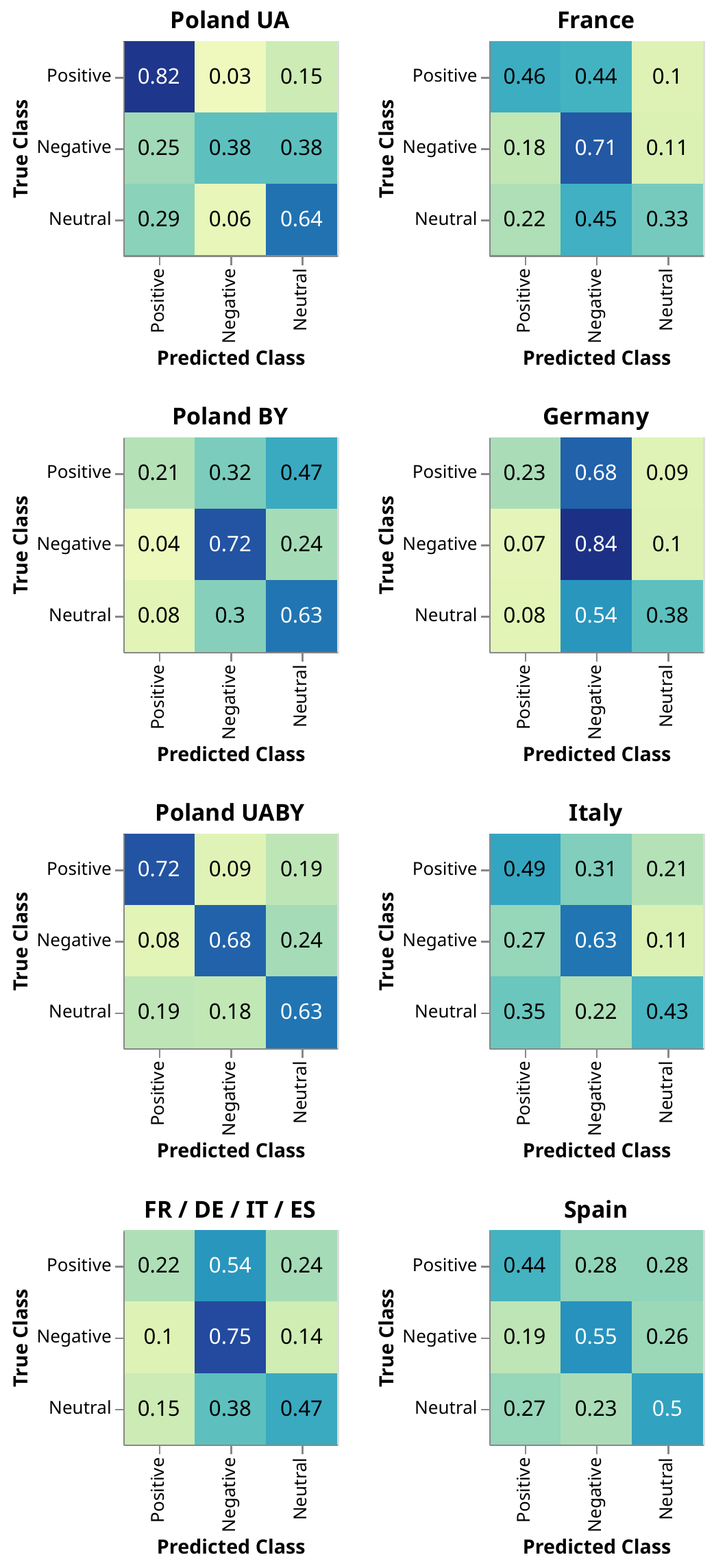}
    \caption{Classification confusion matrices for individual datasets.}
    \label{fig:confusion-multi}
    \Description{The diagonal in the matrices is most pronounced for Poland UABY, France, Italy and Spain.}
\end{figure}

For further insights in the reply stance classifier performances on the individual datasets as described in Section~\ref{subsec:stance-detection-cross-linguality} and Table~\ref{table:classification-performance}, we provide the respective confusion matrices in Figure~\ref{fig:confusion-multi}.

\end{document}